\documentclass[twocolumn,amsmath,amssymb,nofootinbib]{revtex4}

\usepackage{graphicx}

\begin{document}

\title{Analytical \lowercase{$r$}-mode solution with
gravitational radiation reaction force}

\author{\'Oscar J. C. Dias\footnote{Also at \uppercase{C}entro
\uppercase{M}ultidisciplinar de \uppercase{A}strof\'{\i}sica --
\uppercase{CENTRA}.}}

\affiliation{Perimeter Institute for Theoretical Physics, 31
Caroline St.\ N., \\
Waterloo, Ontario N2L 2Y5, Canada \\
E-mail: odias@perimeterinstitute.ca}

\author{Paulo M. S\'a}

\affiliation{Departamento de F\'{\i}sica and Centro
Multidisciplinar de Astrof\'{\i}sica -- CENTRA, \\ Universidade do
Algarve, Campus de Gambelas, 8005-139 Faro, Portugal \\
E-mail: pmsa@ualg.pt}

\begin{abstract}
We present and discuss the analytical $r$-mode solution to the
linearized hydrodynamic equations of a slowly rotating, Newtonian,
barotropic, non-magnetized, perfect-fluid star in which the
gravitational radiation reaction force is present.
\end{abstract}

\maketitle

\section{Introduction}

The $r$-modes, which are pulsation modes of rotating stars that
have the Coriolis force as their restoring force, are driven
unstable by gravitational radiation (GR) \cite{Andersson1998}. In
the frame co-rotating with the star, the energy of the unstable
$r$-mode grows exponentially, $E=E_0 e^{-2t/\tau_{GR}}$, with the
gravitational timescale $\tau_{\rm GR}$ given by
\cite{LindOwenMors1998,AndersKokkSchutz1999}
\begin{eqnarray}
\tau_{\rm GR} = - \left(
    \frac{2^{17}\pi}{5^2\times3^8} \, \frac{G}{c^7} \, \tilde{J}
    \Omega^6      \right)^{-1}\,,
 \label{GRtimescale:LindOwenMors}
\end{eqnarray}
where $G$ is the Newton's constant, $c$ is the velocity of light,
$\Omega$ is the angular velocity of the star and $\tilde{J}\equiv
\int_0^R dr \hat{\rho} r^6$ with $\hat{\rho}$ and $R$ being,
respectively, the density and the surface's radius of the
unperturbed star. For a wide range of relevant temperatures and
angular velocities of newly born, hot, rapidly-rotating stars,
bulk and shear viscosity do not suppress the exponential growth of
the energy of the $r$-mode
\cite{LindOwenMors1998,AndersKokkSchutz1999}.

In the above-mentioned investigations
\cite{LindOwenMors1998,AndersKokkSchutz1999}, the linearized
hydrodynamics equations with the GR force were used to obtain an
expression for the time evolution of the physical energy of the
$r$-mode perturbation, $dE/dt$, from which the gravitational
radiation and viscous timescales were determined. In this work, in
which the main results of Ref.~\cite{DiasSa1} are reported, we
present an explicit expression for the $r$-mode velocity
perturbations that solves the linearized hydrodynamics equations
with the GR force. Our conclusions are that (i) these velocity
perturbations are sinusoidal with the same frequency as the
well-known GR force-free linear $r$-mode solution, (ii) the GR
force drives the $r$-modes unstable with a growth timescale that
agrees with the expression found in
Refs.~\cite{LindOwenMors1998,AndersKokkSchutz1999}, and (iii) the
amplitude of these velocity perturbations is corrected, relatively
to the GR force-free case, by a term of order $\Omega^6$.

\section{Hydrodynamic equations with GR reaction force}

The Newtonian hydrodynamic equations for a uniformly rotating,
barotropic, non-magnetized, perfect-fluid star in the presence of
the gravitational radiation (GR) reaction force are the Euler,
continuity, and Poisson equations given, respectively, by
\begin{eqnarray}
&
\partial_t \vec{v} + ( \vec{v} \cdot \vec{\nabla} ) \vec{v}
 = - \rho^{-1} \vec{\nabla} P - \vec{\nabla} \Phi + \vec{F}^{\rm GR},
&
 \label{Euler}
\\
&
\partial_t \rho + \vec{\nabla} \cdot (\rho \vec{v}) = 0,
&
 \label{Continuity}
\\
&
 \nabla^2 \Phi = 4 \pi G \rho,
&
 \label{Poisson}
\end{eqnarray}
where $\rho$, $P$ and $\vec{v}$ are, respectively, the density,
the pressure and the velocity of the fluid, $\Phi$ is the
Newtonian potential, and the GR reaction force,
\begin{equation}
 \vec{F}^{\rm GR} = - \partial_t \vec{\beta} + \vec{v} \times
 ( \vec{\nabla} \times \vec{\beta} ),
 \label{DefGravForce}
\end{equation}
is assumed to be given by the 3.5 post-Newtonian order expansion
that includes the contribution of the current quadrupole moment,
which is the main responsible for the GR instability that sets in
\cite{Blanchet1997,Rezzolla1999}. In Eq.~(\ref{DefGravForce}),
$\vec{\beta}$ is the gravitational vector potencial whose
components are given by
\begin{equation}
 \beta_i \equiv \frac{16G}{45c^7} \, \epsilon_{ijk} x_j x_q
   S_{k q}^{[5]},
 \label{DefBeta}
\end{equation}
where $S_{ij}(t)$ is the time-varying current quadrupole tensor,
\begin{equation}
 S_{i j}(t) \equiv \int d^3x \epsilon_{kq(i}x_{j)} x_k \rho v_q,
 \label{DefSij}
\end{equation}
$\epsilon_{i j k}$ is the Levi-Civita tensor, $x_i$ is the
Cartesian coordinate of the point at which the tensor is
evaluated, and $S_{i j}^{[n]}(t)$ denotes the $n^{th}$ time
derivative of $S_{i j}$.

\section{Linearized hydrodynamics equations with GR reaction force}
\label{LinearTheory}

The hydrodynamic equations (\ref{Euler})--(\ref{Poisson}) can be
linearized, yielding
\begin{eqnarray}
& \partial_t \delta^{(1)} \! \vec{v} + ( \delta^{(1)} \! \vec{v}
\cdot \vec{\nabla} ) \hat{\vec{v}} + ( \hat{\vec{v}} \cdot
\vec{\nabla} ) \delta^{(1)} \! \vec{v}  \nonumber \\
&= - \vec{\nabla} \delta^{(1)} \! U + \delta^{(1)} \! \vec{F}^{\rm
GR},&
 \label{EulerLinearized}
\\
&
\partial_t \delta^{(1)} \! \rho + \hat{\vec{v}} \cdot \vec{\nabla}
\delta^{(1)} \! \rho + \vec{\nabla} \cdot \left( \hat{\rho}
\delta^{(1)} \! \vec{v} \right) = 0,
 \label{ContinuityLinearized}
&
\\
&
 \nabla^2 \delta^{(1)} \! \Phi = 4\pi G \delta^{(1)} \! \rho,
 \label{PoissonLinearized}
&
\end{eqnarray}
where $\hat{\vec{v}}=\Omega r \sin\theta \vec{e}_{\phi}$ is the
velocity of the unperturbed star, $\hat{\rho}$ its mass density,
$\delta^{(1)} \! Q$ denotes the first-order Eulerian change in a
quantity $Q$ and we have defined $\delta^{(1)} \! U \equiv
\delta^{(1)} \! P/\hat{\rho} + \delta^{(1)} \! \Phi$.

To compute the first-order Eulerian change in the GR force,
$\delta^{(1)} \! \vec{F}^{\rm GR}$, we assume that the GR
force-free $r$-mode velocity perturbations\footnote{In this
article, we will be concerned exclusively with the $l=2$ $r$-mode,
which is the most unstable mode.},
\begin{subequations}
 \label{r-mode:noGRforce}
\begin{eqnarray}
 \delta^{(1)} \! v_r &=& 0,
\\
 \delta^{(1)} \! v_{\theta} &=& -\frac{i}{4} \sqrt{\frac{5}{\pi}}
\frac{\alpha\Omega}{R} r^2
 \sin\theta e^{i(2\phi+\omega t)},
\\
 \delta^{(1)} \! v_{\phi} &=& \frac14 \sqrt{\frac{5}{\pi}}
   \frac{\alpha\Omega}{R}
   r^2 \sin\theta \cos\theta
 e^{i(2\phi+\omega t)},
\end{eqnarray}
\end{subequations}
act as a source for the first-order Eulerian change in the current
quadrupole tensor, $\delta^{(1)} \! S_{ij} $. In the above
expression, $\alpha$ is the amplitude of the $r$-mode and we
assume that
\begin{equation}
\omega = \omega_0 + i \varpi, \label{Def:omegaTotal}
\end{equation}
where the frequency of the $r$-mode, $\omega_0\equiv {\rm
Re}[\omega]$, and the small imaginary part that is related to the
growth timescale of the instability of the mode, $\varpi={\rm
Im}[\omega]<0$, are arbitrary parameters to be determined.

Under the above assumptions, $\delta^{(1)} \! S_{xx}$ is computed
to be
\begin{eqnarray}
\delta^{(1)} \! S_{xx} = -\alpha \Omega \sqrt{\frac{\pi}{5}}\,
 \frac{\tilde{J}}{R} \, e^{-\varpi t} e^{i\omega_0 t},
 \label{deltaS:total}
\end{eqnarray}
where we have neglected the contribution coming from the terms
$\hat{v}_i \delta^{(1)} \! \rho$ (of order $\alpha \Omega^3$) and
retained only the dominant terms $\hat{\rho}\delta^{(1)} \! v_i$
(of order $\alpha \Omega$). Similarly, it is straightforward to
show that the first-order Eulerian change in the other components
of the quadrupole tensor satisfy the relations
\begin{eqnarray}
& &  \delta^{(1)} \! S_{yy} = -\delta^{(1)} \! S_{xx} , \nonumber \\
& &  \delta^{(1)} \! S_{xy} = i\delta^{(1)} \! S_{xx} ,\nonumber \\
& &  \delta^{(1)} \! S_{xz} = \delta^{(1)} \! S_{yz} =
\delta^{(1)} \! S_{zz} = 0.
 \label{deltaSij:all}
\end{eqnarray}

Using Eqs.~(\ref{deltaS:total}) and (\ref{deltaSij:all}), the
first-order Eulerian change in the gravitational vector potential,
$\delta^{(1)} \! \vec{\beta}$, is then computed to be
\begin{subequations}
 \label{beta}
\begin{eqnarray}
\hspace{-5mm} \delta^{(1)} \! \beta_r &=& 0,
\\
\hspace{-5mm} \delta^{(1)} \! \beta_{\theta} &=& -\kappa
\frac{\tilde{J}}{R}
     \alpha \Omega \omega_0^5 r^2
 \sin\theta e^{-\varpi t} e^{i(2\phi+\omega_0 t)},
\\
\hspace{-5mm} \delta^{(1)} \! \beta_{\phi} &=& -i\kappa
\frac{\tilde{J}}{R}
     \alpha \Omega \omega_0^5r^2
 \sin\theta \cos\theta e^{-\varpi t} e^{i(2\phi+\omega_0 t)},
\end{eqnarray}
\end{subequations}
where the constant $\kappa$ sets the strength of the GR reaction
force and is defined as
\begin{equation}
 \kappa \equiv \frac{16}{45} \sqrt{\frac{\pi}{5}} \,\frac{G}{c^7} .
 \label{Def:kappa}
\end{equation}

Finally, taking into account that $\hat{S}_{ij} = 0$, implying
that $\hat{\beta_i} = 0$, the first-order Eulerian change in the
GR force, $\delta^{(1)} \! \vec{F}^{\rm GR}$, is computed to be
\begin{subequations}
 \label{deltaForce:end}
\begin{eqnarray}
\delta^{(1)} \! F_{r}^{\rm GR}  &=&  - 3 i \kappa
\frac{\tilde{J}}{R} \alpha \Omega^2 \omega_0^5 r^2 \sin^2\theta
\cos\theta \nonumber \\ && \times e^{-\varpi t} e^{i(2\phi +
\omega_0 t)},
\\
\delta^{(1)} \! F_{\theta}^{\rm GR} &=& i \kappa
\frac{\tilde{J}}{R} \alpha \Omega \omega_0^5 \left( \omega_0 +
3\Omega \sin^2 \theta \right) r^2 \sin\theta \nonumber
\\ && \times e^{-\varpi t} e^{i(2\phi+\omega_0 t)},
\\
\delta^{(1)} \! F_{\phi}^{\rm GR} &=& -\kappa \frac{\tilde{J}}{R}
\alpha \Omega \omega_0^6  r^2 \sin\theta \cos\theta \nonumber \\
&& \times e^{-\varpi t} e^{i(2\phi+\omega_0 t)}.
\end{eqnarray}
\end{subequations}

\section{The analytical \emph{r}-mode solution with
GR reaction force}

The linearized hydrodynamic equations
(\ref{EulerLinearized})--(\ref{PoissonLinearized}), with the
first-order Eulerian change in the GR force, $\delta^{(1)} \!
\vec{F}^{\rm GR}$, given by Eq.~(\ref{deltaForce:end}), admit the
solution \cite{DiasSa1}
\begin{subequations}
 \label{r-mode:GR}
\begin{eqnarray}
\delta^{(1)} \! v_r &=& 0,
\\
\delta^{(1)} \! v_{\theta} &=& - \frac{i}{2} \alpha \Omega \left[
\frac12 \sqrt{\frac{5}{\pi}} \frac{1}{R} + i \gamma \left( A-1
\right) \Omega^5 \right] r^2 \sin\theta \nonumber \\ & & \times
e^{-\varpi t} e^{i(2\phi+\omega_0 t)},
\\
\delta^{(1)} \! v_{\phi} &=& \frac12 \alpha \Omega \left[ \frac12
\sqrt{\frac{5}{\pi}} \frac{1}{R} + i \gamma \left( A-1 \right)
\Omega^5 \right] r^2 \sin\theta \cos\theta \nonumber \\ & & \times
e^{-\varpi t} e^{i(2\phi+\omega_0 t)},
\end{eqnarray}
\end{subequations}
and
\begin{eqnarray}
\delta^{(1)} \! U &=& \frac{1}{3} \alpha \Omega^2 \left( \frac12
\sqrt{\frac{5}{\pi}} \frac{1}{R} +i \gamma A \Omega^5 \right) r^3
\sin^2\theta \cos\theta \nonumber \\ && \times e^{-\varpi t}
  e^{i(2\phi+\omega_0 t)},
 \label{r-mode:GR:U}
\end{eqnarray}
with $\varpi$ and $\omega_0$ given by
\begin{equation}
\varpi =- \frac{8}{9}\sqrt{\frac{\pi}{5}}\, \gamma R \Omega^6
\quad \mbox{and} \quad \omega_0 = - \frac{4\Omega}{3}.
\end{equation}

The velocity perturbations given by Eq.~(\ref{r-mode:GR}) have a
piece similar to the GR force-free solution, the difference being
the factor $e^{-\varpi t}$ responsible for the exponential growth
of the $r$-mode amplitude due to the presence of a GR reaction
force, and another piece proportional to $\alpha\gamma (A-1)
\Omega^6$, where $A$ is a constant fixed by the choice of initial
data and $\gamma \equiv 1024\kappa\tilde{J}/(81R)$.

The GR force-free limit is obtained when we set the parameter
$\kappa$, defined in Eq.~(\ref{Def:kappa}), equal to zero. In this
limit, $\varpi$ and $\gamma$ also go to zero. Then, from
Eqs.~(\ref{r-mode:GR}) and (\ref{r-mode:GR:U}), we recover the GR
force-free linear $r$-mode solution.

\section{Discussion of the results and future directions}

In this work we have presented the analytical $r$-mode solution to
the linearized Newtonian hydrodynamic equations with the GR
reaction force. The velocity perturbations $\delta^{(1)} \!
\vec{v}$ are proportional to $e^{i(2\phi+\omega_0 t)}$, with
$\omega_0=-4\Omega/3$. Thus, they have the same sinusoidal
behavior and the same frequency $\omega_0$ as the GR force-free
velocity perturbations given by Eq.~(\ref{r-mode:noGRforce}). The
amplitude of the velocity perturbations is proportional to
$\exp\{-\varpi t\}$. Since $\varpi<0$, the GR force induces then
an exponential growth in the $r$-mode amplitude. The e-folding
growth timescale $\tau_{\rm GR}=1/\varpi$ agrees with the GR
timescale (\ref{GRtimescale:LindOwenMors}) found in
Refs.~\cite{LindOwenMors1998,AndersKokkSchutz1999}. The velocity
perturbations $\delta^{(1)} \! \vec{v}$ contain also a piece
proportional to $\alpha \gamma (A-1)\Omega^6$, where $A$ is an
arbitrary constant fixed by the choice of initial data. If we
choose this constant $A$ to be of order unity, then this part of
the solution could be neglected \cite{DiasSa1}.

A quite interesting feature that has emerged from recent
investigations on $r$-modes is the presence of differential
rotation induced by the $r$-mode oscillation in a background star
that is initially uniformly rotating. That differential rotation
drifts of kinematical nature could be induced by $r$-mode
oscillations of the stellar fluid was first suggested in
Ref.~\cite{RezzLambShap1}. The existence of these drifts was
confirmed in numerical simulations of nonlinear $r$-modes carried
out both in general relativistic hydrodynamics
\cite{StergFont2001} and in Newtonian hydrodynamics
\cite{LindTohVal2001}. Differential rotation was also reported in
a model of a thin spherical shell of a rotating incompressible
fluid \cite{LevinUsh2001}. Recently, an analytical solution,
representing differential rotation of $r$-modes that produce large
scale drifts of fluid elements along stellar latitudes, was found
within the nonlinear Newtonian theory up to second order in the
mode amplitude and in the absence of GR reaction \cite{Sa2004}.
This differential rotation plays a relevant role in the nonlinear
evolution of the $r$-mode instability \cite{SaTome2005}. Two
questions could be then naturally raised, namely, is differential
rotation also induced by the gravitational radiation reaction and
does this differential rotation play a relevant role in the
nonlinear evolution of the $r$-mode instability? One of the aims
of the investigation carried out in Ref.~\cite{DiasSa1} is to
initiate a programme that hopefully will allow to answer this
question. The natural continuation of this investigation is then
to try to find an analytical $r$-mode solution of the
\emph{nonlinear} hydrodynamic equations with the GR reaction
force. This work is in progress \cite{DiasSa2}.

\section*{Acknowledgments}
It is a pleasure to thank Kostas Kokkotas and Luciano Rezzolla for
interesting discussions. This work was supported in part by the
\emph{Funda\c c\~ao para a Ci\^encia e a Tecnologia} (FCT),
Portugal. OJCD acknowledges financial support from FCT through
grant SFRH/BPD/2004.

\end{document}